\documentclass[useAMS,usenatbib,usegraphicx]{mn2e}
\usepackage{times}

\def\xmm{{\em XMM-Newton}}
\def\c{{\em Chandra}}

\def\ukirt{{\em UKIRT}}
\def\spitzer{{\em Spitzer}}

\def\iras{{\em IRAS}}
\def\vla{{\em VLA}}
\def\p{$\pm$}
\def\ltsim{\mathrel{\hbox{\rlap{\hbox{\lower4pt\hbox{$\sim$}}}\hbox{$<$}}}}
\def\gtsim{\mathrel{\hbox{\rlap{\hbox{\lower4pt\hbox{$\sim$}}}\hbox{$>$}}}}

\def\Msun{M$_{\odot}$}
\def\micron{$\mu$m}

\def\nh{$N_{\rm H}$}

\def\iraf{{\sc iraf}}
\def\xspec{{\sc xspec}}

\def\cloudy{{\sc cloudy}}
\def\ha{H$\alpha$}
\def\hb{H$\beta$}

\def\oiii{[O{\sc iii}]}
\def\oii{[O{\sc ii}]}

\def\sii{[S{\sc ii}]}

\def\l{$\lambda$}

\def\a96315{A963\_15}
\def\cxouj_a96315{CXOU~J101714.1+390124}

\title{4C~+39.29 -- Extended emission around a powerful Type 2 quasar}

\author[P. Gandhi, A.C. Fabian and C.S. Crawford]
{\parbox[]{6.in} { P. Gandhi$^{1,2\dag}$, A.C. Fabian$^{1}$ and C.S. Crawford$^{1}$\\
\footnotesize
$^{1}$Institute of Astronomy, Madingley Road, Cambridge CB3 0HA\\
$^{2}$European Southern Observatory, Alonso de Cordova 3107, Casilla 19001, Santiago, Chile\\
$^{\dag}$pg@ast.cam.ac.uk\\}}
\date{Accepted 2006 March 31.  Received 2006 March 24; in original form 2006 February 14.}

\begin{document}

\maketitle

\begin{abstract}
We present new X-ray and optical spectroscopy of a Type 2 quasar candidate selected from a follow-up program of hard, optically-dim, serendipitous \c\ sources. The source is confirmed to be a Type 2 quasar at $z=0.536$ with an intrinsic 2--10 keV luminosity $L_{2-10}=5\times 10^{44}$ $h_{0.7}^{-2}$ erg s$^{-1}$, an absorbing column density $N_{\rm H}=8\times 10^{23}$ cm$^{-2}$ and a 
neutral Fe K$\alpha$ line detected by \xmm\ EPIC-MOS1 as well as \c\ ACIS-S.
An extended optical forbidden emission line cloud 
is detected at the same redshift, and at about 15 kpc in projected separation. 
This cloud lies in close proximity to the peak of the compact steep spectrum radio source 4C~+39.29, which has previously been identified with a foreground galaxy in the cluster Abell~963.
We present evidence to show that 4C~+39.29 is associated with the background X-ray Type 2 quasar. The radio luminosity is dominated by lobes with complex structure and the radio core is weak in comparison to narrow-line radio galaxies at the same X-ray luminosity.
The morphology and emission line properties of the extended region are consistent with an on-going jet--cloud interaction. 
4C~+39.29 possesses a combination of high power and high absorbing column density compared with other X-ray Type 2 quasars in the literature.
These observations highlight the efficacy of using X-rays to identify the primary energy source of complex radio sources and distant obscured AGN.

\end{abstract}

\section{Introduction}

Powerful jets and associated lobes and hotspots make radio galaxies and quasars visible to vast distances across the Universe. In crowded fields, however, these extended (and often, distorted) structures can make counterpart identification tricky. The radio source 4C~+39.29 is such a case; it lies in the direction of the richness class 3 cluster of galaxies Abell~963, but the issue of whether the radio source is associated with the cluster, or not, has been unresolved thus far.\footnote{the NASA Extragalactic Database assigns the tentative redshift of $z=0.206$ (of Abell~963) to the radio source.}

The problem can be understood from Fig.~\ref{fig:images}, which shows an archival $i$-band (SDSS filter) image of the field of 4C~+39.29 from the William Herschel Telescope, overlaid with 3.6~cm radio contours extracted from an archival radio observation (described in \S~\ref{sec:radio} below). The peak of the radio emission is seen to be approximately coincident with the optical source marked ``2''. This is an $R\sim18.5$ galaxy at RA=10h17m14.2s Dec=$+39^\circ 01'21.5''$~(J2000). Yet, the field is crowded (there are 6 detections within a radius of 10$''$ of the radio peak above the limiting 3$\sigma$ magnitude of 23.4). Given the comparatively lower resolution of early optical and well as radio (low frequency) observations, it is not surprising that all optical sources marked 2, 3 as well as 4 in Fig.~\ref{fig:images} ({\sl left}) were proposed as possible counterparts to the radio source, though source 2 was usually considered most likely \citep{riley75, wills76, padrielliconway77, owen93, mcmahon02}. 

The advent of superb imaging capabilities in the X-ray domain provides a complementary observational technique to identify the correct counterpart, since X-ray emission is a ubiquitous characteristic of active galactic nuclei (AGN) that are thought to power radio jets and lobes \citep[e.g., ][]{mushotzky93}. \citet{g04} discovered, using \c\ observations (\S~\ref{sec:xray}), an X-ray source in the field of 4C~+39.29, but associated it with the faint, red galaxy ($B\sim 22.8; R\sim 19.8; B-K=6.3$) that lies 4 arcsec to the N-W of the previous identification. This identification [at 10:17:14.1 +39:01:24 (J2000)], marked ``1'' in Fig.~\ref{fig:images}, was possible due to the excellent spatial resolution of \c\ and the presence of a weak emission feature in the X-ray spectrum, the observed energy of which agreed with a photometric redshift estimate of 0.56\p0.1, if identified as Fe~K$\alpha$. This suggested that the X-ray source lay behind Abell~963, not associated with the cluster.

But is there any physical connection between the X-ray and radio sources? 4C~+39.29 has long been known to have at least two radio components separated by approximately 8 arcsec (e.g., \citealt{saikia01, machalskicondon83}), with the southern component being brighter and complex, while the northern component has a fainter, but normal, jet-like morphology ending in a hot spot. These components are seen in Fig.~\ref{fig:images} and, interestingly, seem to be centred on the optical galaxy marked 1 that is associated with the X-ray source, suggesting that this may be the source of radio emission. 

In this paper, we present detailed observations at X-ray, optical and radio wavelengths of the field of 4C~+39.29. We find that the likely counterpart of 4C~+39.29 is source 1, which is a powerful Type 2 quasar obscured by column densities of cold gas approaching $N_{\rm H}\sim 10^{24}$ cm$^{-2}$. We present newly analysed \xmm\ observations to support the Type 2 quasar hypothesis, already suggested in \citet{g04}. New optical spectroscopy of the field uncovers spatially extended emission lines at the redshift of source 1, and provides reinforcement of the counterpart identification. We then discuss different mechanisms that could power the bright emission of the extended radio and forbidden line structures and suggest that a jet--cloud interaction is likely to be on-going. 
Compared to other narrow-line radio quasars, 4C~+39.29 has a weak radio core, and suggests that the relation between compact steep spectrum radio sources and X-ray Type 2 quasars should be further explored. 4C~+39.29 lies in the regime of the brightest and most obscured AGN when compared to other Type 2 quasars. 
We assume $H_0=70$ km s$^{-1}$ Mpc$^{-1}$, $\Omega_{\rm M}=0.3$ and $\Lambda=0.7$, where required.

We note that some initial results consistent with ours, including possible non-association of 4C~+39.29 with the cluster Abell~963, as well as detection of extended line emission, were reported by \citet{lavery93}. To our knowledge, however, no result other than an abstract has been published.

\section{X-ray observations}
\label{sec:xray}
\subsection{{\em \textbf Chandra}}

The X-ray source \cxouj_a96315\ was discovered by \cite{g04} during a serendipitous search for optically-dim sources with hard X-ray spectral count ratios in the fields of medium-deep \c\ ACIS-S observations (typically, observations of relaxed galaxy clusters available at that time; see Table~\ref{tab:xrayobservations} for observation details). The source, designated source 15 in the field of Abell~963 (or A963\_15) was easily identified as having an extremely hard $S/H$ count ratio of 0.08.\footnote{$S/H \equiv \rm{Cts_{(0.5-2 keV)}/Cts_{(2-7 keV)}}$.} Only a faint optical counterpart was visible on Digitized Sky Survey images. Though there was some ambiguity in source identification, the most likely counterpart assigned in cross-correlation with medium-deep \ukirt\ near-infrared images was the source marked 1 in Fig.~\ref{fig:images}. The extracted X-ray spectrum itself gave an indication of the source redshift, showing a peak at about 4.1~keV, possibly implying $z\sim 0.5$ if the feature was Fe K$\alpha$. The line was detected at 90 per cent significance, with the data allowing a large equivalent width.

The line was seen to be superposed on a hard X-ray continuum, consistent either with an absorbed power-law with $N_{\rm H}\sim 10^{24}$ cm$^{-2}$ or with a reflection model (with an additional foreground absorber). In either case, the intrinsic, absorption-corrected X-ray luminosity of the source was well above typical Seyfert luminosities, making this a Type 2 quasar. The ACIS spectrum is shown in Fig.~\ref{fig:xspec}, and we refer the reader to \citet{g04} for full details on the \c\ analysis.

\subsection{{\em \textbf XMM-Newton}}

We retrieved an archival observation of the field of Abell~963 taken with \xmm\ (see Table~\ref{tab:xrayobservations}). Standard data reduction tasks of creating filtered event lists and extraction of images/spectra were carried out using {\sc sas}~v~6.5.0, with the latest calibration information. Fortunately, the observation was not affected by flaring. 

Unfortunately, \cxouj_a96315\ lies in the gap between CCDs 4 and 5 of the EPIC-pn camera, the one with the largest effective area. Furthermore, the source lies at a projected distance of 2.5~arcmin from the core of Abell~963, and is affected by the extended intra-cluster foreground, especially in the soft band.

\subsubsection{EPIC-MOS1}
\label{sec:mos1}
The EPIC-MOS1 data are shown in Fig.~\ref{fig:xspec} along with the \c\ ACIS-S spectrum. 
Due to the bright extended cluster emission, several different circular and elliptical apertures $\sim 20-40$ arcsec in size were tried for source count extraction in each instrument. Larger regions representative of the average noise at the position of the source were used for background extraction. The MOS1 observation showed approximately $100$ net counts above 2 keV. Lower energies are significantly contaminated by the cluster foreground, as well as some possibly extended source emission, and were ignored in the modeling below. For completeness, we note that a \lq leaky\rq\ absorber, with a covering fraction of \lq holes\rq\ of 0.2 per cent, could fit the data between 1--2 keV, but any soft emission is likely to be of a different origin, as discussed below in \S~\ref{sec:extendedxrays}.

Each spectral channel was binned to have a minimum of 10 counts. The flat (hard) slope of the data is apparent above 3 keV. We first fit a redshifted, absorbed power-law model to the MOS1 spectrum alone (using $z=0.536$; see \S~\ref{sec:optspec} below). Fig.~\ref{fig:xspec} clearly shows the excess residuals at an observed energy of $\sim 4.1$ keV in the MOS1 spectrum as well, exactly where a redshifted Fe K$\alpha$ line in the source frame is expected to lie. The rest energy of the line is inconsistent with H-like (6.7 keV) or He-like (6.96 keV) Iron; we thus included a neutral line with fixed rest-energy of 6.4~keV and a narrow width of 10~eV in the model. Galactic absorption due to a fixed column density of $N_{\rm H}=1.4 \times 10^{20}$ cm$^{-2}$ (based on observation of {\sc Hi}; \citealt{stark92}) was also included in all fits. 

Fixing the photon-index slope to $\Gamma=1.9$ results in a very large column density of $N_{\rm H}=1.2_{-0.5}^{+0.8} \times 10^{24}$ cm$^{-2}$ local to the source (errors corresponding to 90 per cent confidence for one interesting parameter using $\chi^2$ statistics), consistent with the source being Compton-thick. The Fe line, with a rest-frame equivalent width (EW) of 900 eV, is significant only at 90 per cent (via the f-test). The intrinsic, absorption-corrected 2--10 keV luminosity is $10^{45}$ erg s$^{-1}$. In the small counts regime, the variance of parameter values is better characterized by the C-statistic, which does not, however, provide an estimate of the goodness-of-fit \citep{xspec}. Using this results in only a marginally reduced best-fit absorption as well as luminosity, well within the error range above. 

Fitting the ACIS and MOS1 spectra together gives a good fit with the above model. Using $\chi^2$ statistics and letting the normalization of each instrument vary freely, the goodness-of-fit is $\chi^2/\nu = 11.2/12$, where $\nu$ are the number of degrees of freedom. The fit is worsened by $\Delta \chi^2 = 8.5$ when the Fe line is frozen out and removed from the fit. This implies a significance level of 97 per cent for the line, which is a stronger result than for either instrument alone. The rest-frame EW itself is not well constrained, and has a value of 0.58$_{-.48}^{+.82}$~keV (approximate 68 per cent re-sampling errors returned by \xspec).

The \c\ and \xmm\ observations were taken 11 months apart, and the normalizations of the two power-law models agree at the 68 per cent level (with flux $F_{2-10}=1.4_{-0.9}^{+0.6} \times 10^{-13}$ erg s$^{-1}$ cm$^{-2}$). Thus, we tried another fit under the assumption of no variability in the physical parameters between the two datasets, i.e. the relevant normalizations and intrinsic absorption for both sets were varied together. With this fit, we obtain a lower 
$N_{\rm H}=8.4_{-2}^{+3} \times 10^{23}$ cm$^{-2}$ and $L_{2-10}=5\times 10^{44}$ erg s$^{-1}$ (90 per cent C-statistic), but the significance of the Fe line is increased to greater than 99 per cent.

The presence of a strong Fe line could also be indicative of strong reflection \citep[e.g., ][]{iwasawa05}. We tried a fit of the {\sc pexrav} model of \citet{pexrav} with an e-folding energy of 300~keV and a $2\pi$ pure reflection model with solar abundance. As for the ACIS spectrum before, we found that the MOS1 spectrum required an additional absorber (of $N_{\rm H}=7_{-4}^{+7} \times 10^{23}$ cm$^{-2}$ if local to the source), and the intrinsic luminosity is inferred to be $L_{2-10}=2.6\times 10^{44} f^{-1}$ erg s$^{-1}$, where $f$ is the reflection albedo. Fitting the MOS1 and ACIS spectra together, with normalizations let free, the absorbed {\sc pexrav} model has a goodness-of-fit $\chi^2/\nu = 11.0/12$, and is indistinguishable from the absorbed power-law model above with the present data.

\subsubsection{EPIC-MOS2}

The MOS2 spectrum of \cxouj_a96315\ (not shown) 
contained about 10 per cent fewer net counts than MOS1. Fitting an absorbed power-law model to only the MOS2 data still allows for high intrinsic absorption of $N_{\rm H}=5.4_{-3}^{+6} \times 10^{23}$ cm$^{-2}$ (C-statistic). There is, however, no evidence for a redshifted Fe K$\alpha$ line; the weak upper limit to the rest-frame equivalent-width that we derive is $\sim 450$ eV. Fitting together the MOS2 data with MOS1 and ACIS decreases the significance of the Fe line slightly to 95 per cent as compared to 97 per cent measured in the previous section. The difference with respect to the EPIC-MOS1 is presumably due to small number statistics combined with the high cluster foreground, but a deeper observation is required to confirm this. 

\subsection{Extended X-ray emission}
\label{sec:extendedxrays}

\begin{figure*}
  \begin{center}
    \fbox{\includegraphics[angle=0,width=8cm]{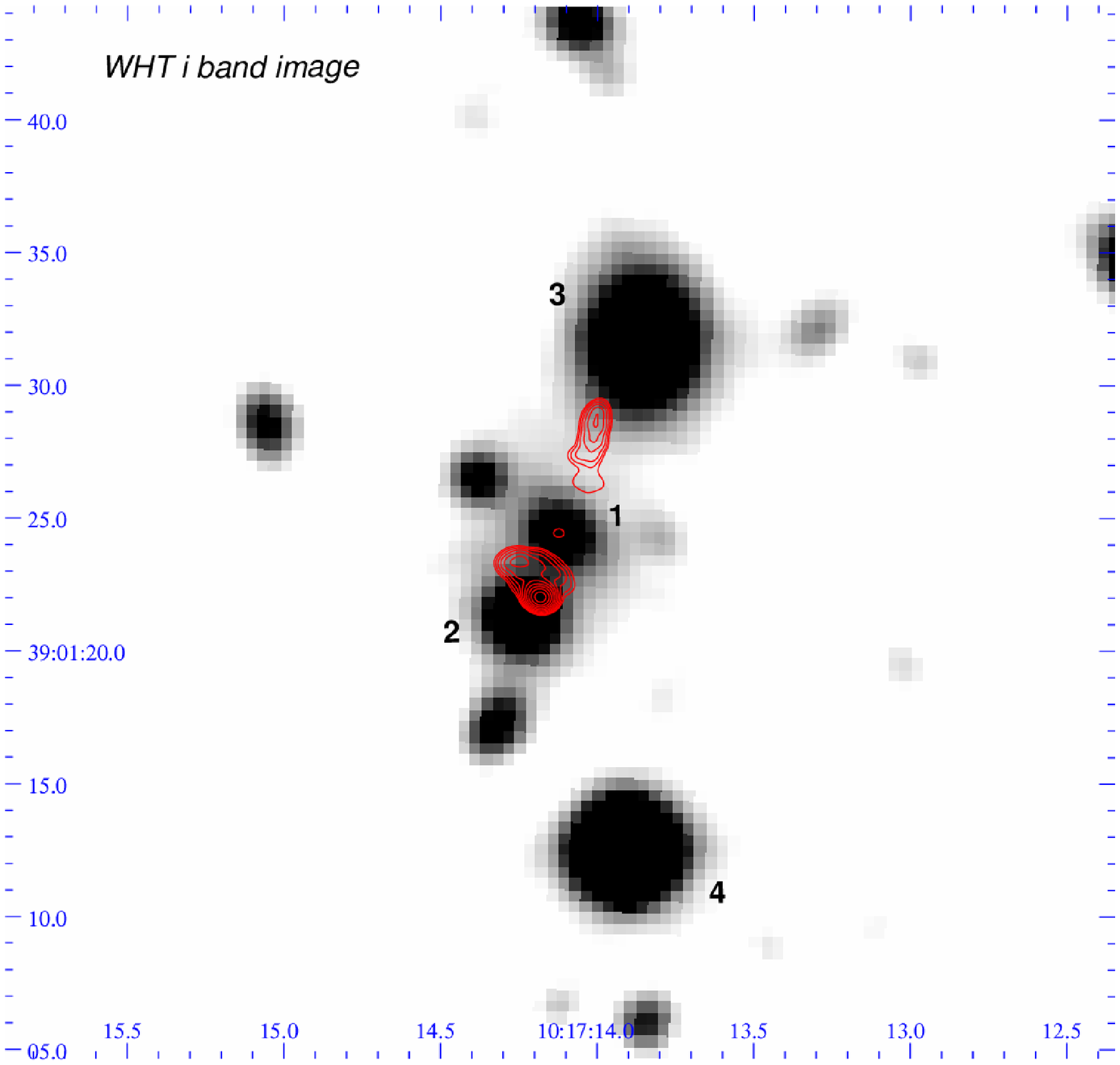}}
    \fbox{\includegraphics[angle=0,width=8cm]{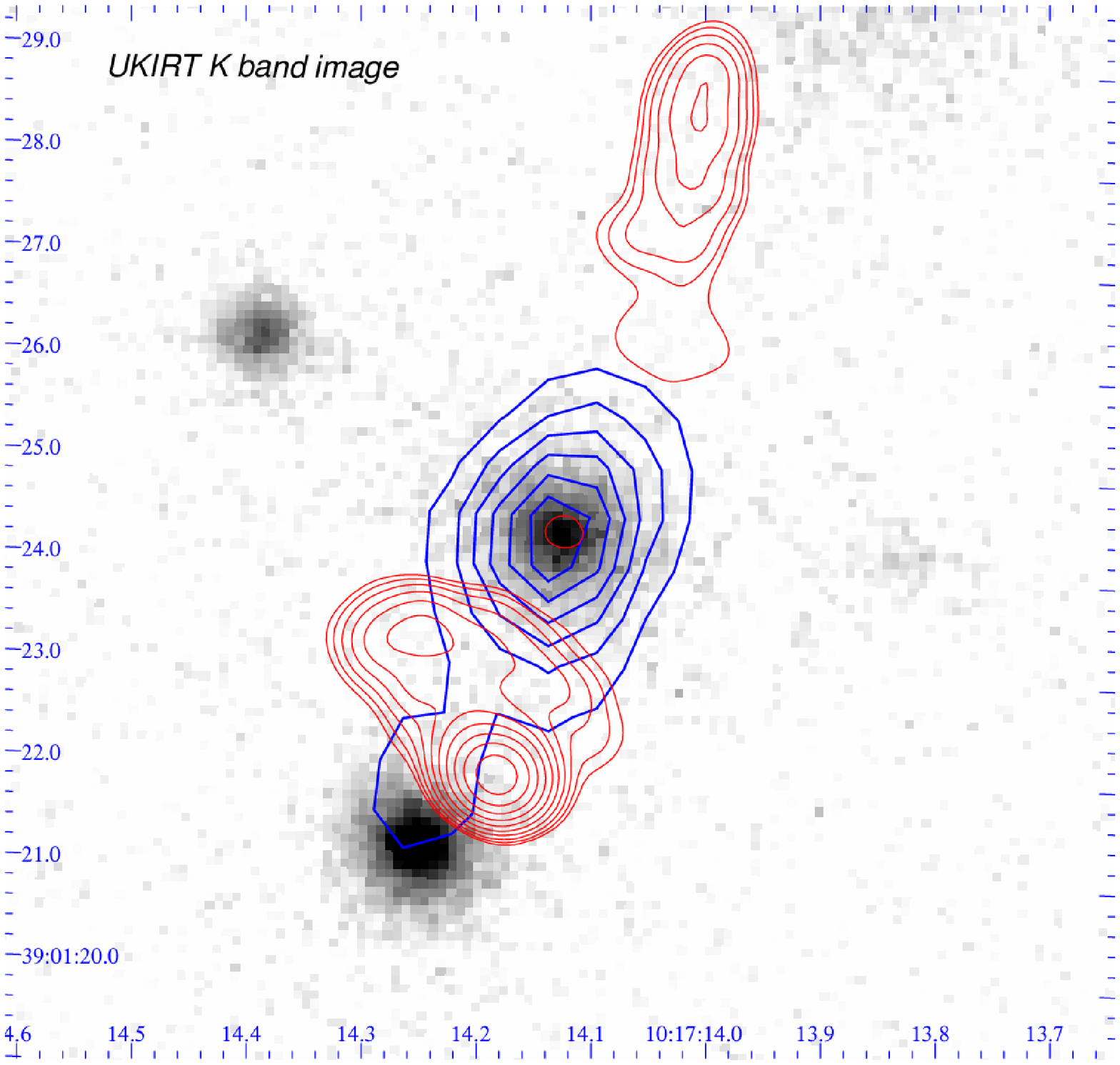}}
    \caption{{\sl (Left)} {\em WHT} $i$-band image of the field ($\sim 35$ arcsec on a side) of 4C~+39.29, with radio \vla\ 3.6 cm contours superposed in red. North is to the top and East is to the left. Sexagesimal coordinates (J2000) are marked along the edges. The optical counterpart usually associated with 4C~+39.29 is source 2, though sources 3 and 4 have also been considered in the literature. In this paper, we present evidence that source 1 is the correct counterpart.
{\sl (Right)} A better sampled {\em UKIRT} $K$-band zoom-in ($\sim 10$ arcsec on a side) of the region around source 1, showing the radio contours again in red, in addition to \c\ 0.5--7 keV contours in blue. 
The X-ray contours are linearly spaced with the faintest one at $\approx 1$ ct pix$^{-1}$ above the local background, while the radio contours follow a square-root progression from 0.5 mJy beam$^{-1}$ upward.
} \label{fig:images} 
  \end{center}
\end{figure*}

The X-ray emission is marginally extended 
along the direction of the radio axis (Fig.~\ref{fig:images}). 
At the location of \cxouj_a96315\ (about 2.5~arcmin S--E of the aim-point of \c) the off-axis point-spread-function (PSF) is also expected to be elongated, so we checked the source parameters returned by the {\sc wavdetect} detection algorithm \citep{wavdetect}, using the full range of $\sqrt{2}$ wavelet scales. The {\sc psfratio} of the source is found to be 1.7. Clearly extended sources should have {\sc psfratio}$>>$1, so evidence for extension is marginal. 

At the faintest level, however, there is an adjacent clump of soft X-rays about 5 counts above the background to the S--E (Fig.~\ref{fig:images}) almost coincident with the position of source 2. This was excluded from the ACIS spectrum, but could contribute to low-energy ($< 2$ keV) residuals seen in the \xmm\ spectrum where it is unresolved. This emission is unlikely to be from any active nucleus in source 2, given its soft X-ray nature, and the fact that the optical spectrum of source 2 (discussed below) is quiescent. If associated with source 1, it could be thermal emission from shocked gas or Inverse Compton emission even though there is not an exact coincidence of the radio emission with the X-ray clump \citep[e.g., ][]{carilli02}. 

\begin{figure*}
  \begin{center}
\includegraphics[angle=270,width=13cm]{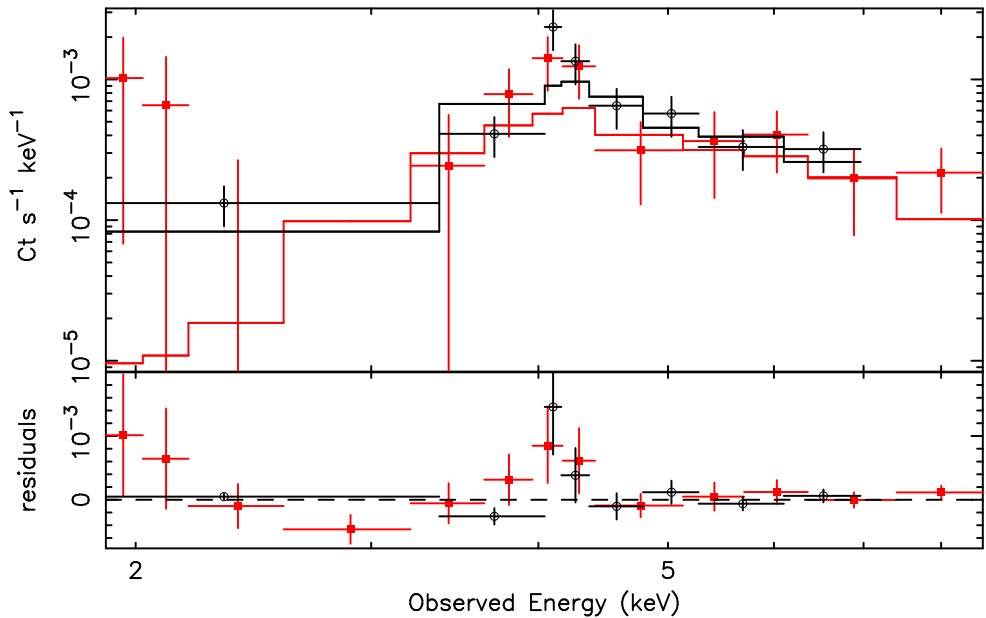}
  \caption{X-ray spectra of \cxouj_a96315\ extracted from \xmm\ EPIC MOS1 (red filled squares and error bars) and \c\ ACIS-S data (black circles and error bars). Best-fit obscured power-law model lines are shown with the same colours as above. Residuals to the fits in the bottom part of the plot show an excess at $\sim$ 4.1 keV in the MOS1 and ACIS spectra consistent with the position of a redshifted Fe K$\alpha$ line. The excess at low energies (below 2 keV) in the MOS1 spectrum is from residuals of the background subtraction due to the projected proximity of the source to the cluster Abell~963, and some possible extended emission.} \label{fig:xspec} \end{center}
\end{figure*}

\begin{table*}
  \begin{tabular}{llcccc}
    Observatory    &    Instrument     &  Obs ID     & Exposure Time  & Obs Date    &  Principal\\
                   &                   & /Seq no.    &      s         &             &  Investigator\\
    \hline          
  \c               &  ACIS-S           &  800079     &    36290       & Oct 11 2000 &  S.W. Allen\\
  \xmm             &  EPIC-MOS         &  0084230701 &    26510       & Nov 02 2001 &  J.-P. Kneib\\
   $''$            &   EPIC-pn         &     $''$    &    20021       &    $''$     &      $''$\\
    \hline
  \end{tabular}
  \caption{Log of X-ray observations analysed.\label{tab:xrayobservations}}
\end{table*}

\section{Optical, Near-infrared Photometry}
\label{sec:optphot}

Broad band optical magnitudes in $UBR$ and $i$ were measured by \citet{g04} by extracting and calibrating archival images from various telescope archives. Combined with near-infrared $JHK$ photometry obtained at {\em UKIRT}, a photometric redshift of $z_{\rm phot}=0.56\pm 0.1$ for source 1 was derived. A 500 Myr Bruzual-Charlot starburst model reddened with $A_V=0.9$ mags of extinction gave a good fit to the broad-band colours. The implied absolute magnitude in the $B$-band is --22, about 5 times more luminous than the magnitude--redshift relation of red field galaxies found by \citet[][ after correcting for differences in the assumed cosmology]{lilly95}. The $K$-magnitude of $16.5 \pm 0.1$ is consistent with the $K-z$ relation for massive radio ellipticals of \citet{eales93}. 

\section{Optical Spectroscopy}
\label{sec:optspec}

An optical spectrum of source 1 was obtained with the ALFOSC instrument at the Nordic Optical Telescope. The observations were carried out in service mode on the night of 2005 Jan 20. The grism used gave a resolution of $\sim 500$ at 6000~\AA. The best range for spectral extraction was restricted to $\sim$4000--8000~\AA\ due to vignetting at the blue end and bright, blended night sky emission at the red end. Conditions during the observations were clear with optical seeing $\sim 0.7-1.3$ arcsec. The useful exposure time was 4000~s (2000~s of additional data were discarded due to poor seeing). 

The raw data were bias-subtracted, flat-fielded, dispersion-corrected and flux calibrated using standard \iraf\ routines. Continuum from source 1 was found to be weak, so a one-dimensional spectrum was extracted from a slit 11 pixels ($\sim$2~arcsec) in width centred on the single brightest emission line detected (see below). The flux density in the source continuum agrees with that inferred from photometry to better than a factor of two (after slit loss corrections).

\begin{figure*}
  \begin{center}
    \includegraphics[angle=90,width=16cm]{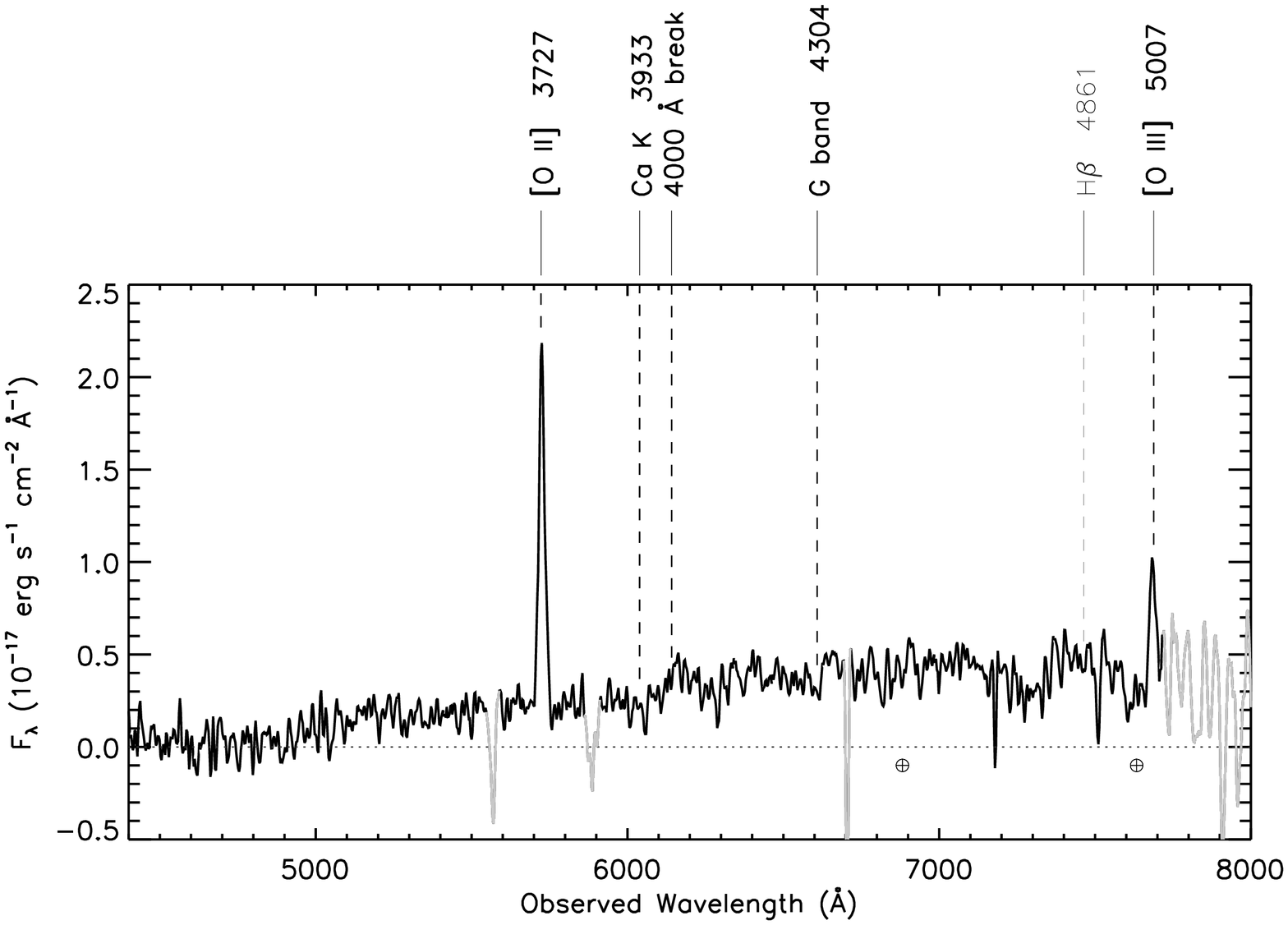}
    \caption{Optical spectrum of source 1 obtained with the Nordic Optical Telescope, smoothed with a moving box average of size 3 pixels for display. Parts of the spectrum significantly affected by sky subtraction residuals are plotted in light grey. Identified features (which all imply $z=0.536$) are labelled in bold; the position of undetected \hb\ is marked in light font. The positions of strong telluric absorption at $\sim$ 6880 \AA\ and 7600 \AA\ are marked by circled crosses.} \label{fig:notspec} 
  \end{center}
\end{figure*}

The final spectrum is shown in Fig.~\ref{fig:notspec}; it shows a very red continuum and we detect two emission lines, that can be identified as redshifted \oii\l3727 and \oiii\l5007 \AA\ at $z=0.536$. Emission line characteristics, measured assuming a single gaussian profile, are listed in Table~\ref{tab:oxygenlines}. In addition, there are absorption features, including weak, redshifted Ca interstellar absorption and the 4000~\AA\ break at $\sim$6100\AA, beyond which the continuum level increases by a factor of about two. The limiting 3$\sigma$ flux at the expected position of the undetected \hb\ line is $1.4 \times 10^{-16}$ erg s$^{-1}$ cm$^{-2}$. 

The strong and narrow forbidden emission lines are consistent with those expected from a Type 2 QSO, hosted by an early-type host galaxy with the red continuum seen \citep[e.g., ][]{norman02}. Compared to other Type 2 quasars with spectra covering a similar rest-frame wavelength range, 4C~+39.29 shows a higher \oii\l3727:\oiii\l5007 ratio of $\approx 2.7$ (whereas \citealt{dellaceca03} and \citealt{caccianiga04} find this ratio to be less than one). This may result from a low ionization parameter, as is characteristic in the nuclei of many central cluster galaxies \citep[e.g., ][]{crawford99}. Further analysis and conclusions regarding the nature of the excitation are deferred for now, however, given that only two emission lines are detected, and one of them (\oiii\l5007) is weak and possibly affected by telluric absorption (Fig.~\ref{fig:notspec}).

\begin{figure*}
  \begin{center}
    \fbox{\includegraphics[angle=0,width=16cm]{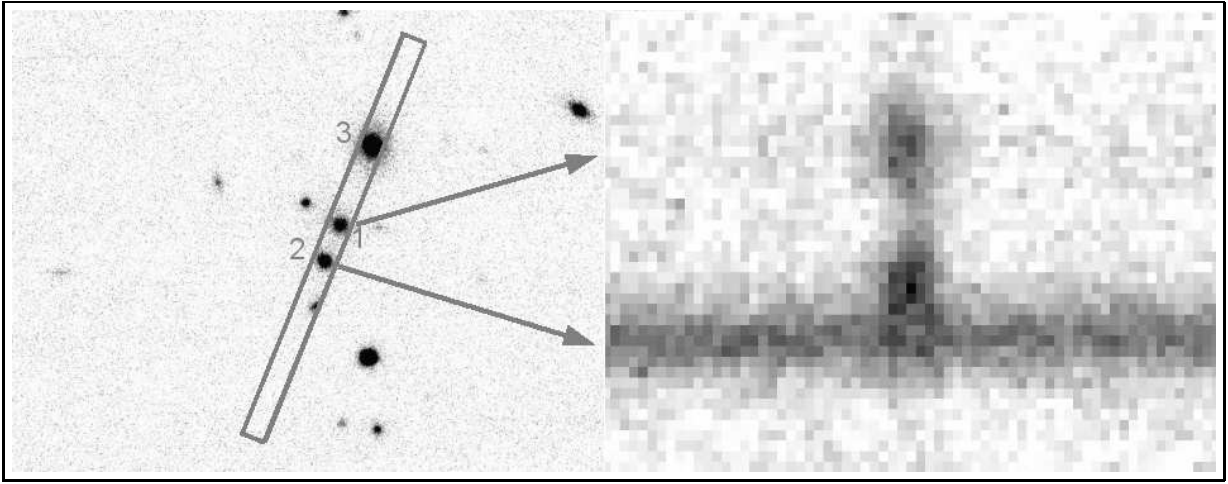}}
    \caption{A \ukirt\ $K$-band image of the field of 4C~+39.29 ({\sl left}) showing the position of the optical spectroscopic slit. ({\sl right}) The two-dimensional spectrum image of the redshifted \oii~\l~3727~\AA\ region in the spectrum of source 1 obtained with the Nordic Optical Telescope. The bright emission line (on very faint continuum) at the top is redshifted \oii. Dispersion direction is from left to right. The bright continuum seen about 16 pixels ($\sim$3~arcsec) below this emission line is the spectrum of source 2. Note the \oii\ line emission between the two spectra, that we associate with spatially extended line emission of source 1.} \label{fig:2dspec} 
  \end{center}
\end{figure*}

\begin{figure*}
  \begin{center}
    \includegraphics[angle=90,width=17cm]{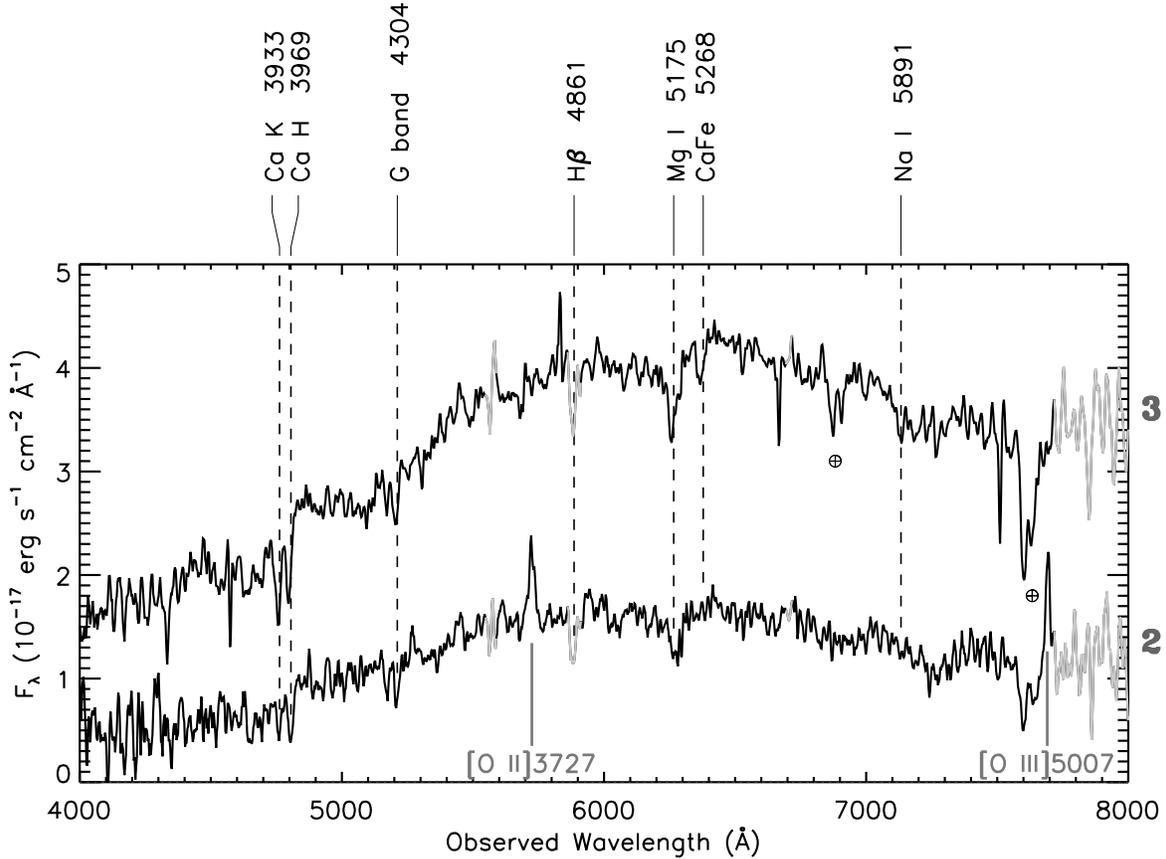}
    \caption{Optical spectra of source 2 ({\em bottom}) at $z=0.210$ and source 3 ({\em top}; offset upwards by $1\times 10^{-17}$ units on the y-axis for clarity of plotting) at $z=0.209$. The emission lines due to forbidden Oxygen transitions in the extended emission region of source 1 at $z=0.536$ were included in the extraction aperture of source 2, and are labelled below the spectrum (see text). See caption of Fig.~\ref{fig:notspec} for other details of the plot.} \label{fig:neighbourspec} 
  \end{center}
\end{figure*}

\begin{table*}
  \begin{tabular}{lcccccr}
    Emission line    & Obs $\lambda$   &     $z$    &  Rest FWHM         &  Line Flux                    &  Line luminosity         &  W$_\lambda$\\
                     &      (\AA)      &            &  (km s$^{-1}$)     &  (10$^{-16}$ erg s$^{-1}$ cm$^{-2}$) &  (10$^{41}$ erg s$^{-1}$)     & (\AA)\\
         (1)         &       (2)       &     (3)    &      (4)           &     (5)                              &           (6)                 &  (7) \\
    \hline
    \oii\l3727        &    5725.2                &    0.5361  &     899$_{-86}^{+87}$ &    3.59$_{-0.51}^{+0.55}$  &  4.1                &    165$_{-38}^{+50}$ \\
    \oiii\l5007       &    7685.0                &    0.5348  &   719$_{-306}^{+392}$ &    1.35$_{-0.78}^{+1.31}$  &  1.5                &    40$_{-25}^{+49}$  \\
    \oii\l3727$_E$    &    5726.5                &    0.5365  &     624$_{-87}^{+88}$ &    2.65$_{-0.51}^{+0.56}$  &  3.0                &     --   \\
    \oiii\l5007$_E$   &    7690.4                &    0.5359  &   683$_{-244}^{+251}$ &    2.05$_{-0.95}^{+1.28}$  &  2.3                &     --   \\
    \hline
  \end{tabular}
  \caption{Observed optical emission line characteristics as measured in the Nordic Optical spectrum. Lines with a subscript $E$ refer to the extended emission regions (\S~\ref{sec:extended}). Errors on line parameters are for a $\Delta \chi^2=2.7$, corresponding to a 90 per cent confidence interval for a single parameter of interest. Column 4 lists the rest-frame full-width at half-maximum of the lines, corrected for instrumental broadening. The final column (W$_\lambda$) lists the observed equivalent widths of the lines. Since no continuum is visible at the position of the extended emission lines, no equivalent widths were measured. 
\label{tab:oxygenlines}}
\end{table*}

\subsection{Spectroscopy of objects in the field}

The spectroscopic slit was placed at a position angle of --24 degrees (East of North) in order to include objects neighbouring source 1 (Fig.~\ref{fig:2dspec}). 
The nearest source on the slit is source 2, about 3.5~arcsec SSE of source 1, and its spectrum is shown in Fig.~\ref{fig:neighbourspec}. Absorption lines typical of am early-type galaxy at $z=0.210$ place the source within the large scale structure associated with the cluster Abell~963 (the implied recession-velocity relative to the cluster is $\sim$1000 km~s$^{-1}$). Two emission lines are also clearly visible in the $\sim 2''$ extraction aperture, and these are discussed in the next section.

Finally, the bright source 3 approximately 7~arcsec NNW of source 1 shows an early-type galaxy spectrum at $z=0.209$ (Fig.~\ref{fig:neighbourspec}), implying that it is also associated with Abell~963.

\subsection{Extended line emission associated with Source 1}
\label{sec:extended}

In Fig.~\ref{fig:2dspec}, there is a clear detection of emission straddling the two horizontal spectra, at the same position on the dispersion axis as the bright \oii~\l~3727 emission line of source 1. Similar emission straddling the two objects is observed at the position of redshifted \oiii~\l~5007 (not shown). The early-type galaxy spectrum of the neighbour (at $z=0.21$) includes these superposed emission lines (Fig.~\ref{fig:neighbourspec}). This is likely to be extended line emitting gas associated with source 1. 
The angular separation between the peak of the \oii~\l~3727 line in source 1 and the peak of the extended emission is $\sim 2.6$ arcsec, corresponds to a projected spatial separation of approximately 17~kpc at $z=0.536$. 

In order to measure the flux of the extended line emission, the sky emission lines and background in the two-dimensional spectrum (Fig.~\ref{fig:2dspec}) were  first subtracted at each pixel of the dispersion axis. Next, the continuum of source 2 was removed by computing a median spatial profile of the source from pieces of continuum free of the extended emission. This profile was subtracted from the above image, resulting in a spectrum of the extended emission free of contamination by the continuum of source 2. This was extracted and calibrated in a manner similar to the spectra of the other objects, using a fixed aperture of 5 pixels (0.95~arcsec) on either side of the peak of the extended emission lines (although no extended continuum is observed, the extraction was straightforward because there is very little spectral curvature on the ALFOSC CCD). The measured extended line parameters are listed in Table~\ref{tab:oxygenlines}, showing that the extended emission is almost as powerful as that seen in source 1 itself. This is discussed further in \S~\ref{sec:discussion}. The limiting flux at the position of redshifted \hb\ is $\sim 6\times 10^{-17}$ erg s$^{-1}$ cm$^{-2}$.

\section{Archival Radio Observations}
\label{sec:radio}

The radio source 4C~+39.29 has, for some time, been known to have a compact and asymmetric morphology (e.g., \citealt{machalskicondon83}). Integrated fluxes at several frequencies smaller than 5~GHz ($\equiv$ 6~cm) are tabulated in the NASA Extragalactic Database (NED). With a spectral index of $\alpha = 0.7\ (S_\nu\propto \nu^{-\alpha})$, it is classified as a compact steep spectrum source \citep[CSS; e.g., ][]{saikia01}. 

We retrieved high-frequency radio observations (NRAO project ID AO111 from 1993 Jan 17) with the Very Large Array (\vla) in the \lq A\rq\ configuration at 2~cm and 3.6~cm ($\equiv$ 15~GHz and 8.5~GHz respectively) from the NRAO archive\footnote{http://archive.nrao.edu/archive/e2earchive.jsp}. These data were analysed using standard routines within the Astronomical Image Processing System ({\sc aips}) package. The source 3C~286 and the compact source 0923+392 were used as primary and secondary calibrators respectively.  

The 3.6~cm image of the source is shown in Fig.~\ref{fig:images}, and fluxes of individual radio components are listed in Table~\ref{tab:radiofluxes}. Image resolution is proportional to frequency, and is about 0.2~arcsec at 3.6~cm. At least two radio components are discerned, aligned approximately along the axis separating sources 1 and 2. The north-western component has a jet-like morphology ending in a point-like hot spot. The southern component, on the other hand, is more complex, with at least two sub-component lobe-like structures on either side of the \lq jet\rq\ axis. The southern-most hotspot is the dominant radio component in the field, and contains at least 60 per cent of the integrated flux at 3.6~cm. This hotspot is itself resolved into two further components in the highest frequency image at 15~GHz (not shown; also noted by \citealt{lavery93}). Using iterative self-calibration to clean the images reveals the presence of a very faint, inverted-spectrum core at both \vla\ frequencies (see Fig.~\ref{fig:images}). The radio coordinates of this point source are 10h17m14.12s +39$^{\circ}$01$'$24$''$.4 (J2000), and the astrometry of our images at all other wavelengths (optical, near-IR, X-ray) has been tied to this position. 

The position of the extended optical forbidden line emission (Fig.~\ref{fig:2dspec}) overlaps with the fainter part of the radio emission seen in the southern lobe. Given that our spectroscopic slit did not cover the exact position of the peak of the southern radio hotspot, it is possible that the radio and extended optical emission are even more closely associated, as discussed in the next section.

\begin{table}
\begin{center}
 \begin{tabular}{lcr}
 Component               &  Flux (3.6cm)       &   Flux (2 cm)\\
                         &    mJy              &       mJy\\
\hline
Total                    &     277             &      143 \\
Southern lobe + hotspots &     236             &      132$^\dag$ \\
Southern hotspot peak    &   $150^*$           &      54$^*$ \\
Northern lobe + hotspot  &      31             &      26\\
Core$^\ddag$             &      0.6            &      1.5\\
\hline
 \end{tabular}
~\par
\end{center}
$^*$ mJy beam$^{-1}$;\, \, \, $^\dag$ At 2 cm, the southern hotspot splits into two, with fluxes of 70 and 30 mJy each.;\, \, \, $^\ddag$ The core flux density is uncertain by a factor of $\approx 2$.
\caption{\vla\ flux densities integrated over various components.\label{tab:radiofluxes}}
\end{table}

\begin{figure*}
  \begin{center}
    \includegraphics[angle=90,width=16cm]{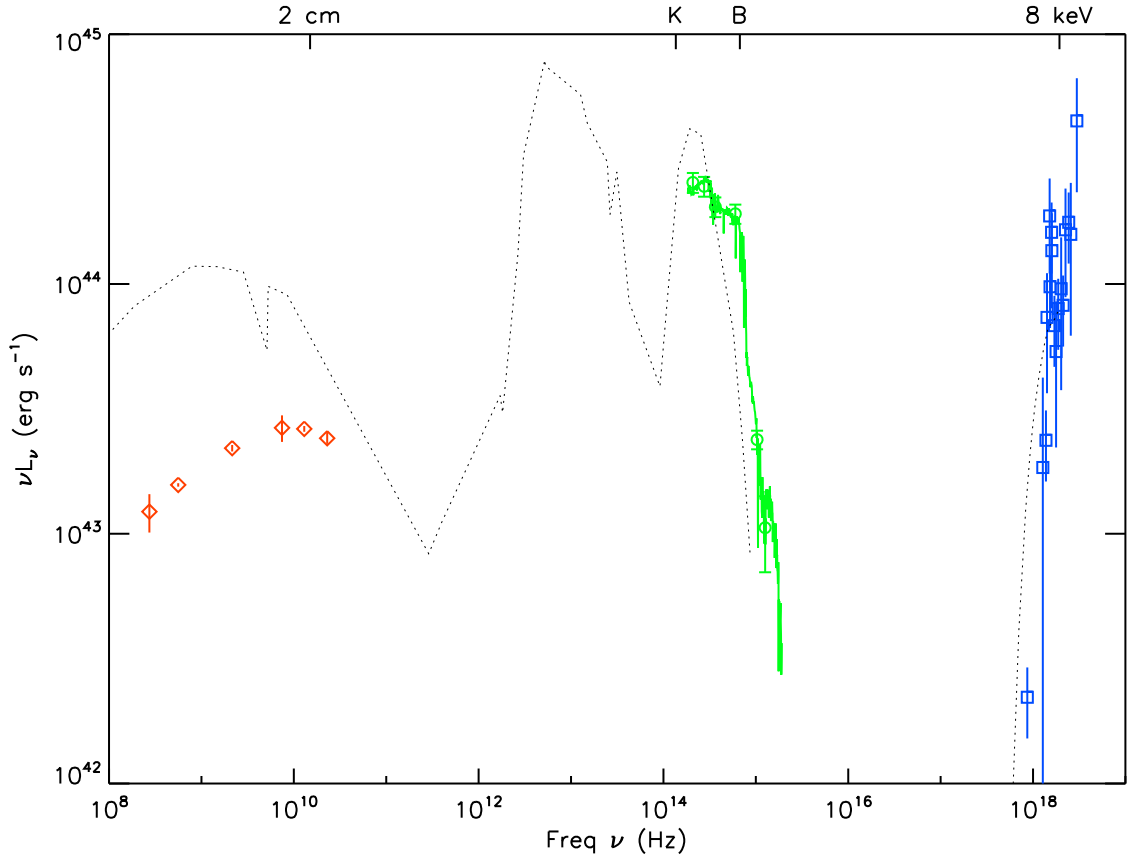}
    \caption{The radio to X-ray spectral energy distribution (SED) of 4C~+39.29, assuming that source 1 is the correct counterpart. The spectrum, plotted in $\nu L_{\nu}$ units, is shown in the source rest-frame at $z=0.536$. The X-ray emission (blue squares) is from the obscured Type 2 quasar, the optical:near-IR emission from the host galaxy (green circles; with the best-fit Bruzual-Charlot SED overplotted), and the radio emission (red diamonds) is from the lobes, predominantly the southern-most hotspot. For comparison, the SED of the FR-II radio galaxy Cygnus A is shown as the dotted line, normalized to the SED of 4C~+39.29 at 8 keV. This SED shows representative radio to ultraviolet values taken from the NED Extragalactic Database, i.e., includes extra-nuclear emission as well. The X-ray part of the SED of Cygnus A represents the best absorbed power-law model of \citet{young02}.
} \label{fig:sed} 
  \end{center}
\end{figure*}

\section{Discussion}
\label{sec:discussion}

\subsection{Source 1 as the counterpart of 4C~+39.29}
\label{sec:sourceonecounterpart}

Previous association of the bright radio hotspot with source 2 (Fig.~\ref{fig:images}) was due to the low resolution radio/optical imaging available at that time. On the other hand, our observation that the two main radio components (the NW and SE lobes/hotspots) are centred approximately on source 1, where a weak core is also discerned at high frequencies, indicates that this is the correct counterpart of 4C~+39.29. The detection of extended emission line gas at the redshift of source 1 along the same axis as the radio lobes, and coincident with the southern radio lobe, is strong evidence to support this hypothesis. Finally, the detection of obscured X-ray emission from source 1 is completely consistent with the hypothesis that this is a powerful radio galaxy driving oppositely-directed radio jets approximately oriented on the plane of the sky. No strong X-ray emission is detected from source 2; nor are optical emission lines seen at the source redshift. Combined with the astrometric offset between the radio lobe and source 2, there is little doubt that this is not the correct counterpart.

\subsubsection{Spectral Energy Distribution}

We thus collate the X-ray (\cxouj_a96315) , optical, near-infrared (source 1) and radio (4C~+39.29) data in Fig.~\ref{fig:sed}, showing the total, rest-frame spectral energy distribution (SED). 
The SED of the powerful radio galaxy Cygnus A is a good overall comparison to 4C~+39.29. Representative integrated radio to ultraviolet fluxes of Cygnus A are compiled from the NED Extragalactic Database. The X-ray part of the Cygnus A SED is the best-fit absorbed power-law nuclear model of \citet[][ Cygnus A is also highly obscured with \nh $\approx 2\times 10^{23}$ cm$^{-2}$]{young02}.

The luminosity and colours of the host galaxy of source 1 are consistent with those of a powerful early-type galaxy (see \S~\ref{sec:optphot}), while any optical light from the nucleus itself is highly extincted by dust associated with the obscuring gas. Assuming a typical optical:X-ray spectral index for the nucleus, $\alpha_{\rm ox}=-1.4$ ($L_\nu \propto \nu^{\alpha_{\rm ox}}$; \citealt{zamorani81}), we can convert the de-absorbed, monochromatic 2 keV X-ray luminosity to that at 2500\AA\ to estimate the extinction. Using the absorbed power-law model for the X-ray emission (\S~\ref{sec:mos1}), we find ($\nu L_{\nu}$)$_{\rm 2500 \AA}^{\rm nuclear}=3.5\times 10^{45}$ erg s$^{-1}$, a factor of 230 higher than the observed (host-galaxy) monochromatic luminosity. Assuming an AGN contribution to 2500\AA\ even as high as 50 per cent, the implied extinction at 2500\AA\ is 7 mags. This is likely to be a very conservative lower-limit. Assuming instead an optical extinction to neutral gas density ratio $A_V:N_{\rm H}$ value lower than the Galactic value by a factor of 10 \citep{maiolino01}, an $A_V\sim 50$ mags is implied from the absorbing X-ray column.

The radio power of 4C~+39.29 is due to the lobes, and is dominated by the southern hotspot. With an integrated, rest-frame power at 178-MHz of $5\times 10^{26}$ W Hz$^{-1}$ sr$^{-1}$, 4C~+39.29 lies well above the dividing luminosity identified by \citet{fanaroff_riley} for their two classes of sources. This is consistent with the lobe-dominated morphology typical of FR II galaxies. Neither of the radio jets themselves are clearly detected, presumably because very little Doppler-boosting is expected for jets directed at large angles to the line-of-sight. The radio core is extremely weak with a rest-frame 13~GHz radio core power ($\nu L_\nu$) of $\sim 5\times 10^{40}$ erg s$^{-1}$, again consistent with the absence of beaming. The ratio between the integrated flux densities of the southern and northern lobes is about a factor of $\approx 2$ if the southern hotspot is excluded (and a factor of $\approx 7$ if included). Such differences have been attributed to density asymmetries in the clumpy media surrounding radio galaxies \citep[e.g., ][]{mccarthy91}.

\subsubsection{Is 4C~+39.29 Compton-thick?}

\citet{bassani99} have explored the possibility of identifying Compton-thick sources by comparing the equivalent width (EW) of the fluorescent Fe K$\alpha$ line with the ratio $T=F_{2-10}/F_{\rm [OIII]}$, with the de-reddened \oiii\ \l 5007 flux from the narrow line region (NLR) assumed to be a measure of the isotropic source luminosity. For 4C~+39.29, we have EW~$\approx 600$ eV and $T_{\rm obs}=630\pm 400$ (error is from propagation of 90 per cent confidence intervals). But in the absence of a measurement of the reddening affecting the \oiii\ line, a determination of the intrinsic value of $T$ is difficult. Using the $L_{{\rm 151\ MHz}} - L_{\rm [O\sc{III}]}$ relation found by \citet{grimes04} for radio galaxies, a crude estimate of the \oiii\ flux decrement of 13 (and up to 400, by using the scatter in their relation) is inferred. We note that obscuring gas associated with the NLR dust causing this decrement could be the same excess absorbing material required by the X-ray reflection-dominated model (\S~\ref{sec:mos1}). Correcting for this reddening places 4C~+39.29 in the regime of highly obscured (\nh~$>4\times 10^{23}$ cm$^{-2}$) sources in Fig.~1 of \citet{bassani99}. Deeper and longer wavelength optical spectra covering \ha\ and \hb\ (to measure the Balmer decrement) will provide better constraints.

\subsection{The nature of the extended emission}
\label{sec:extendedemission}

The observed radio and emission line properties of 4C~+39.29 are similar to those observed for many radio galaxies. For instance, the bright extended emission line region (EELR) lies on the side of the nucleus with the smaller projected distance to the radio lobe \citep{mccarthy91}, and is closely aligned with the radio axis \citep{mccarthy87}. No continuum is detected at the position of the EELR, however, and star-formation is unlikely to be a significant excitation mechanism. 

EELRs are also seen in the $10-100$ kpc scale environments of QSOs \citep[e.g., ][]{stocktonmackenty87} and are thought to be excited by the hard quasar continuum, to which they usually have an unobscured line-of-sight along the radio axis. Such photo-ionization modeling of line intensity ratios around radio-loud quasars has shown the extended gas to be at high pressure or density if ionized by the quasar continuum, suggesting the presence of larger, confining media
 \citep[e.g., ][]{crawfordfabian89}. We followed \citet{fabian87} and attempted to model the EELR line intensities in 4C~+39.29 by using the photo-ionization code \cloudy\ \citep[version 05.07, ][]{cloudy0507}. Simultaneously satisfying the observed \oii:\oiii\ ratio of $1.3 \pm 0.7$ with the limit of \oiii:\hb$>$3 proved to be difficult, however (since the former ratio requires a low ionization parameter [$\xi$], while the latter limit pushes $\xi$ in the other direction). Marginally consistent solutions were obtained with a total Hydrogen column density of 10$^{1.6-2.2}$ cm$^{-3}$, and a correspondingly high gas pressure of $P=nT\approx 1.8\times 10^6$ cm$^{-3}$ K.

\subsubsection{A possible jet--cloud interaction}

On the other hand, the good spatial coincidence between the extended optical emission line gas and the brightest radio lobe of 4C~+39.29 suggests that the radio jet itself (unseen) could provide the ionization source for the optical emission lines. Such jet--cloud interactions have been observed in several radio galaxies (e.g., \citealt{clark97, crawfordvanderriest97}). The mechanical energy of the jet can directly shock-ionize the gas, or provide a high-energy photon field that acts as an ionizing precursor to the shock. Facts supporting this interpretation are:

\begin{enumerate}
\item {\em Radio--Optical alignment:} There is a close projected alignment of the EELR with a hotspot along the radio axis, as would be expected in a physical interaction; 
\item {\em Kinematics:} The linewidths of the observed forbidden lines in the EELR are large: upto 600--700 km s$^{-1}$. Increased linewidths are expected following turbulent motions of an interaction, but not in passive photo-ionization by the AGN \citep{clark97}. Since our spectroscopic slit did not cover the position of the radio peak itself, we cannot rule out the presence of even larger motions in adjacent regions just outside our spectroscopic slit. This slit positioning may also be why we do not see significant radial velocity shifts (Table~\ref{tab:oxygenlines});
\item {\em Morphology:} The southern lobe of radio emission has a complex structure (Fig.~\ref{fig:images}) that may have prompted \citet{lavery93} to consider gravitational lensing as a possibility. Interaction of a jet with a dense, clumpy medium is also likely to produce a complex morphology (e.g., \citealt{pedelty89b}), as plasma and matter diffuse outwards following ram pressure and shock related relaxation; 
\item {\em Radio polarization:} Whereas multi-frequency detections of the polarized intensity of 4C~+39.29 have not been published, \citet{conway77} state an upper-limit of fractional polarization at 6~cm to be less than 2.7 per cent. The source was not resolved in their observation, but the southern radio lobe is undoubtedly the dominant component. Such an absence of polarization is expected due to the high rotation measure of the dense, irregular medium of the EELR, if the synchrotron emitting radio plasma is indeed interacting with it 
\citep[e.g., ][]{clark97}. Full mapping of the multi-wavelength depolarization is needed in order to confirm this; 
\item {\em Shock excitation:} 
Spectral signatures expected from fast, radiative shocks propagating through typical narrow-line AGN clouds have been investigated by \citet{dopitasutherland95}. They find that the key parameters controlling the emission line ratios are the shock velocity and magnetic parameter (that depends on the strength of the magnetic field and the gas density in the cloud). The inclusion of \lq precursor\rq\ ionization due to hard radiation generated in the wake of the post-shock, cooling plasma that diffuses outwards is crucial to explain differences seen in Seyfert and LINER galaxies. We find that the observed \oii:\oiii\ ratio and \oiii/\hb\ limit are consistent with a \lq shock+precursor\rq\ model with shock velocity exceeding $\sim 250$ km s$^{-1}$ at least, and with a finite ($>0\ \mu$G cm$^{3/2}$) magnetic parameter (See Fig.~1 of \citeauthor{dopitasutherland95})

Alternatively, the jet power could simply act to compress the gas, effectively decreasing its ionization parameter, consistent with the large \oii/\oiii\ ratio observed (Table~\ref{tab:oxygenlines}). In this case, shock compression/ionization could be any additional effect, above that expected from AGN photo-ionization. \newline

\noindent
Given that we detect only two extended emission lines along a single slit position, the comparative importance of shock or radiative ionization cannot be further quantified. While the data favour shock ionization occurring in a jet--cloud interaction, pure photo-ionization cannot be completely ruled out from the line intensities alone. We note, however, that \citet{best00} found from a study of $z\sim 1$ 3CR radio sources that sources with a small linear extent ($\la 120$ kpc) show emission line properties consistent with being excited by radiative shocks. Several of the observed properties of 4C~+39.29, including the high power in the \oii\l3727 line and the low ionization state, are easily accommodated within the overall picture inferred by \citeauthor{best00} Sensitive spectra covering the ultraviolet or longer wavelengths, with emission lines such as the \sii\l\l6717,6731 doublet, are needed. These will better constrain the physical conditions of the EELR (densities, temperatures etc.) itself. 

\end{enumerate}

\subsubsection{Lensing of the southern lobe by source 2?}

\citet{lavery93} mentioned the possibility of gravitational lensing to explain the complex radio morphology of the southern lobe, though no details other than their unpublished abstract are available to us. 
In any case, it is pertinent to check if lensing due to the foreground source 2 could boost the luminosity of the background hotspot/EELR. A rough prediction of the expected lensing parameters has been made based on the observed configuration. The putative lens, source 2, is an early-type galaxy with an absolute $K$ magnitude of $-23.3$, making it a $0.6\ K^*$ system, using the near-infrared luminosity function of field galaxies by \citet{gardner97}. The angular separation between the lens (source 2) and the peak of the radio emission is $\approx 1$~arcsec, corresponding to a physical radius of 3.5~kpc the redshift of source 2. Assuming an approximate enclosed mass within this region of $10^{11}$~\Msun\ \citep[cf. ][]{broadhurstlehar95} for a point lens model, the Einstein radius is predicted to be $\approx 0.7$~arcsec, for a background source (the radio lobe/hotspot of 4C~+39.29) that lies at $z=0.536$. Given the crudeness of this estimate as well as possible systematic astrometric uncertainties, the fact that the predicted Einstein radius is close to the observed lens--image separation implies that simple boosting of the flux from the radio lobe of source 1 due to lensing by the foreground source 2 cannot be ruled out. No obvious multiple images of the southern radio hot spot are detected on the other side of source 2, however. Furthermore, a search for faint multiple images of the \oii\ emission line around the spectrum of source 2 in Fig.~\ref{fig:2dspec} did not reveal any \citep[see, e.g., ][]{willis05b}. To summarize, whereas the observed configuration may be consistent with a gravitational lensing solution, the present data do not provide strong {\em positive} evidence for this.

\subsection{Comparison to other Type 2 quasars and radio galaxies}

\begin{figure}
  \begin{center}
    \includegraphics[angle=90,width=8.5cm]{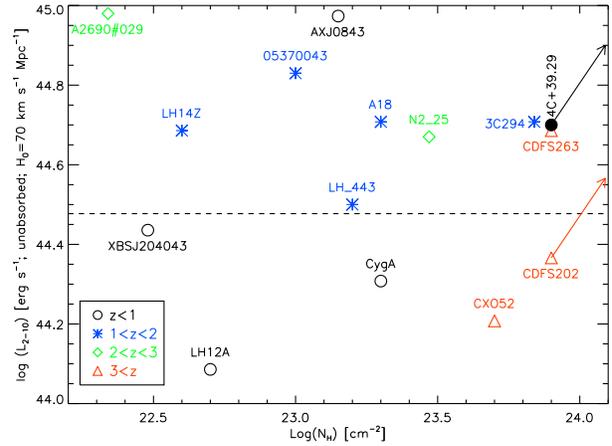}
    \caption{Comparison of intrinsic luminosities and fitted obscuring column densities for X-ray selected Type 2 quasars from the literature. All luminosities have been converted to the 2--10 keV band using a power-law (with $\Gamma=1.8$, if unspecified), and corrected for differing Hubble constants. The plot is not meant to be a complete compilation, but includes some of the most luminous, securely-identified (i.e., with a spectroscopic redshift, and usually a good X-ray spectrum) obscured quasars from a number of recent X-ray surveys. Only power-law transmission model parameters have been considered, since corrections in the case of Compton-thick obscuration are uncertain, though arrows indicate (schematically) the general direction of correction in the case of Compton-thick obscuration if discussed in the literature. This then excludes some clear reflection-dominated quasars such as IRAS~09104+4109 \citep{iras09mainref}. Some luminous sources with $L_{2-10}>>10^{45}$ erg s$^{-1}$ are off the scale, but these usually have columns log(\nh)~$<<23$ \citep[e.g., ][]{severgnini06}. Plotted sources are as follows:-- CDFS263, CDFS202 \citep{mainieri05, norman02, tozzi06}; CXO52 \citep{stern02}; N2\_25 \citep{willott03}; A18 \citep[Abell~2390 field, ][]{c02, g04}; AXJ0843 \citep{dellaceca03}; LH443, LH12A and LH14Z \citep[Lockman Hole,][]{sturm06, mainieri02, mateos05}; 05370043, A2690\#029 \citep[HELLAS2XMM,][]{perola04, maiolino06}; XBSJ204043 \citep{caccianiga04}. Also overplotted for comparison are the radio galaxies 3C~294 \citep{3c294mainref} and Cygnus A \citep{young02}. The dashed line is the canonical Seyfert/quasar divide of $L_{2-10}=3\times 10^{44}$ erg s$^{-1}$.} 
\label{fig:compareqso2s}
  \end{center}
\end{figure}

Fitting a simple power-law at $z=0$ to the X-ray spectra shown in Fig.~\ref{fig:xspec} results in a best-fit photon-index of $\gamma\approx -2\ [N_{\rm E} \propto E^{-\gamma}]$. This is, of course, not physically viable, but can be used for comparison with other samples of powerful, obscured AGN. For instance, the above power-law is flatter than all the Compton-thick candidate spectra presented by \citet{tozzi06} in a study of the X-ray sources identified in an ultra-deep exposure of the Chandra Deep Field South (their Fig.~A.21). Furthermore, with $L_{2-10}>5\times 10^{44}$ erg s$^{-1}$, 4C~+39.29 is more luminous than all their Compton-thin Type 2 quasars at least. In fact, 4C~+39.29 possesses a combination of high power as well as high obscuration when compared to many other Type 2 quasars in the literature (Fig.~\ref{fig:compareqso2s}). It is at least comparable to CDFS263, the Type 2 QSO detected in the sub-mm by \citet{mainieri05} and to the powerful radio galaxy 3C~294 \citep{3c294mainref}, both of which are likely to have high black hole masses ($\ga 10^8$ \Msun). Unlike 4C~+39.29 however, neither shows an obvious Fe K$\alpha$ line. Note that the canonical Seyfert/quasar divide is drawn at $3\times 10^{44}$ erg s$^{-1}$ because this is the 2--10 keV luminosity expected for a typical Bolometric luminosity of 10$^{46}$ erg s$^{-1}$ with a Bolometric correction of $\approx 30$ \citep{elvis94}; many potential Type 2 quasars would not fulfil this criterion.
Given the difficulty of identifying distant, heavily obscured AGN with columns approaching the Compton-thick limit (see e.g., \citealt{fabianwilmancrawford02} for predictions; \citealt{tozzi06} \& \citealt{iwasawa05} for observations), 4C~+39.29 is a good example of such an object that is bright enough to be amenable to detailed study. 

While the integrated radio emission of 4C~+39.29 is typical of radio-loud objects (comparable, say, to the Type 2 radio-loud QSO identified by \citealt{dellaceca03}), the nucleus of 4C~+39.29 is radio-weak. The X-ray to infrared part of the SED (due to the nucleus and host galaxy; Fig.~\ref{fig:sed}) matches well with objects used in the literature as template comparisons for radio-quiet Type 2 quasars (e.g., NGC~6240; \citealt{iwasawa99}).
Using the monochromatic radio core power (\S~\ref{sec:sourceonecounterpart}) and the inverted spectral-index found for the core ($\alpha\approx -1.5$), we deduce $L_{\rm 1-10\ GHz}^{\rm nuc}=10^{40}$ erg s$^{-1}$. For a range of different spectral indices upto $\alpha=0.7$ (similar to that of the lobes), we find $L_{\rm 1-10\ GHz}^{\rm nuc}<3\times 10^{41}$ erg s$^{-1}$. Thus, $L_{\rm X}^{\rm nuc}$/$L_{\rm 1-10\ GHz}^{\rm nuc}> 1600$, where $L_{\rm X}$ is the 2-10 keV de-absorbed, rest-frame luminosity of 4C~+39.29. \citet{carilli02} found that this ratio is about $\sim 500$ for the narrow-line radio quasar PKS~1138--262 as well as for Cygnus A. Thus, the core radio:X-ray power of 4C~+39.29 is lower than that of typical narrow-line radio quasars. Finally, we also note that the compact scale of the radio lobes ($\sim 30$ kpc, typical of CSS sources) as well as the lack of powerful diffuse X-ray emission implies that 4C~+39.29 is not located at the centre of a massive cluster extending out to Mpc scales.

Type 2 quasars, as well as CSS radio sources are thought to represent the early stages in the evolution of AGN \citep[e.g.,][]{f99, fanti95}. AGN could form from accretion under high column densities and sky covering fractions of matter that feed as well as obscure the nucleus. Feedback from the growing AGN will eventually begin to scatter the optically thick clouds surrounding it, allowing radiation and matter to punch through the obscuring \lq embryo\rq. The radio source will then be \lq born\rq\ only when hot plasma begins to expand outwards, and CSS radio sources will then represent transition objects, in which the obscuring matter around an AGN begins to dissipate, or to settle into the canonical shape of a torus. The radio lobes will eventually expand outwards and grow to lengths of hundreds of kpc (and also drop in luminosity; \citealt{blundellrawlings99}). In the case of 4C~+39.29, interaction with an asymmetric, dense and clumpy medium is probably responsible for one radio hotspot being particularly bright. A targeted search for fainter, compact radio lobes around X-ray selected Type 2 quasars (and, vice-versa, an X-ray study of known CSS radio sources) is needed to test this evolutionary picture. Current multi-wavelength studies of Type 2 quasars (e.g., the optically-selected sample of \citealt{zakamska04_II}) do not possess the resolution to probe the compact scales involved.

\section{Conclusions}

4C~+39.29 is a bright, steep spectrum radio source with a complex morphology. In this paper, we have presented evidence to suggest that the X-ray source \cxouj_a96315\ powers 4C~+39.29. We have also confirmed the photometric redshift estimate of \citet{g04}, implying that \cxouj_a96315\ is a Type 2 quasar with powerful, narrow optical emission lines, as well as intrinsically-luminous and obscured X-ray emission, at $z=0.536$. 

The newly-analysed \xmm\ data in this paper support the obscured quasar hypothesis, and allow for the presence of an Fe K$\alpha$ line on a flat continuum. This increases the significance of the line compared to the previous \c\ measurement, but due to photon statistics, high foreground due to cluster thermal emission and unfortunate source location within a gap in the EPIC pn field-of-view, the improvement is not as high as hoped. Assuming a simple transmission model, the intrinsic X-ray luminosity of the source is $L_{2-10}\sim 5\times 10^{44}$ erg s$^{-1}$, under an obscuring column of $N_{\rm H}\sim 10^{24}$ cm$^{-2}$ approaching the Compton-thick limit. The source is thus an interesting example of the handful of distant Type 2 quasars currently well-studied in X-rays. 4C~+39.29 is also one of the few distant AGN for which an indication of the correct redshift was obtained first from the X-ray data \citep[cf., ][]{braito05}.

Strong optical forbidden line emission at the redshift of the Type 2 quasar is detected at a projected distance of at least 15~kpc from the optical early-type galaxy counterpart. The near coincidence of this EELR with the peak of the radio emission (that lies in the southern hotspot) is strong evidence that 4C~+39.29 is a radio galaxy at this higher redshift. Compared to other powerful radio galaxies, the core is radio-weak. Several diagnostics including the complex morphology, the absence of polarized radio emission as well as large measured linewidths are consistent with the interpretation of a jet--cloud interaction occurring in the EELR. Predictions of line excitation due to the ionization field of a fast radiative shock easily match the observed line ratios. The compact nature of the source would also favour shock ionization, in comparison with studies of 3CR radio galaxies \citep{best00}. Pure photo-ionization by the unabsorbed quasar nucleus can marginally fit the line intensity ratios and limits, and implies that the EELR gas must be highly pressurized. Observations of other spectroscopic lines are required in order to study the excitation mechanism and measure any reddening in the EELR and as well as the narrow line region.

One important part of the SED of 4C~+39.29 (Fig.~\ref{fig:sed}) that is missing is the infrared regime, where its peak is likely to lie. The present X-ray data cannot differentiate between the transmission model above, and a reflection model with additional foreground absorption that would imply an intrinsic luminosity higher by the inverse of the scattering albedo ($f$) into the line-of-sight. But interesting constraints can be placed on $f$ from the far-infrared power, since this is an estimate of the bolometric luminosity ($L_{\rm Bol}$) of the source \citep[e.g., ][]{wilman03_hyperluminous}. The \iras\ Sky Survey Atlas\footnote{http://irsa.ipac.caltech.edu/} only provides a limit of $4\times 10^{46}$ erg s$^{-1}$ to the far-infrared monochromatic power at $z=0.536$ (for a limiting point-source flux of 1.2~Jy at 100 \micron). Assuming this to be an approximate upper-limit to $L_{\rm Bol}$, and using a bolometric correction of 30 for the intrinsic 2--10 keV quasar luminosity \citep{elvis94}, we find $L_{\rm 2-10} = 2.6\times 10^{44}f^{-1} < 0.033\times L_{\rm Bol}$ for the pure reflection model (\S~\ref{sec:mos1}). This gives a lower-limit to $f$ of 0.2, i.e., the covering fraction of any material scattering X-rays into the line-of-sight must be 20 per cent at the (unobserved) nucleus, at least. Detection of the source in the far-infrared regime, say with \spitzer, will provide further, important constraints.

\section{Acknowledgements}

PG is supported by an ESO (European Southern Observatory) Fellowship. ACF and CSC acknowledge the Royal Society for support. PG profited from discussions on the radio analysis with Frederik T. Rantakyr\"{o} and Annalisa Celotti, and on the X-ray analysis with Kazushi Iwasawa. The help of Robert J.H. Dunn with the radio analysis is especially appreciated. Roderick M. Johnstone provided extensive help with computer-related issues. We thank the referee, R. Della Ceca, for comments that improved the paper. 

We thank the staff of the Nordic Optical Telescope for carrying out service mode observations. The data presented here have been taken using ALFOSC, which is owned by the Instituto de Astrofisica de Andalucia (IAA) and operated at the Nordic Optical Telescope under agreement between IAA and the NBIfAFG of the Astronomical Observatory of Copenhagen. This work has made use of archival \xmm\ data. This research has made use of the NASA/IPAC Extragalactic Database (NED) which is operated by the Jet Propulsion Laboratory, California Institute of Technology, under contract with the National Aeronautics and Space Administration.

\bibliographystyle{mnras}                       
\bibliography{gandhi4c3929}

\begin{thebibliography}{}

\bibitem[\protect\citeauthoryear{{Arnaud}}{{Arnaud}}{1996}]{xspec}
{Arnaud} K.~A., 1996, in ASP Conf. Ser. 101: Astronomical Data Analysis
  Software and Systems V, eds. George H. Jacoby and Jeannette Barnes, Vol.~5,
  p.~17

\bibitem[\protect\citeauthoryear{{Bassani} et~al.}{{Bassani}
  et~al.}{1999}]{bassani99}
{Bassani} L., {Dadina} M., {Maiolino} R., {Salvati} M., {Risaliti} G., {della
  Ceca} R., {Matt} G.,  {Zamorani} G., 1999, \apjs, 121, 473

\bibitem[\protect\citeauthoryear{{Best}, {R{\"o}ttgering}, \& {Longair}}{{Best}
  et~al.}{2000}]{best00}
{Best} P.~N., {R{\"o}ttgering} H.~J.~A.,  {Longair} M.~S., 2000, \mnras, 311,
  23

\bibitem[\protect\citeauthoryear{{Blundell} \& {Rawlings}}{{Blundell} \&
  {Rawlings}}{1999}]{blundellrawlings99}
{Blundell} K.~M.,  {Rawlings} S., 1999, \nat, 399, 330

\bibitem[\protect\citeauthoryear{{Braito} et~al.}{{Braito}
  et~al.}{2005}]{braito05}
{Braito} V., {Maccacaro} T., {Caccianiga} A., {Severgnini} P.,  {Della Ceca}
  R., 2005, \apjl, 621, L97

\bibitem[\protect\citeauthoryear{{Broadhurst} \& {Lehar}}{{Broadhurst} \&
  {Lehar}}{1995}]{broadhurstlehar95}
{Broadhurst} T.,  {Lehar} J., 1995, \apjl, 450, L41

\bibitem[\protect\citeauthoryear{{Caccianiga} et~al.}{{Caccianiga}
  et~al.}{2004}]{caccianiga04}
{Caccianiga} A. et~al., 2004, \aap, 416, 901

\bibitem[\protect\citeauthoryear{{Carilli} et~al.}{{Carilli}
  et~al.}{2002}]{carilli02}
{Carilli} C.~L., {Harris} D.~E., {Pentericci} L., {R{\"o}ttgering} H.~J.~A.,
  {Miley} G.~K., {Kurk} J.~D.,  {van Breugel} W., 2002, \apj, 567, 781

\bibitem[\protect\citeauthoryear{{Clark} et~al.}{{Clark}
  et~al.}{1997}]{clark97}
{Clark} N.~E., {Tadhunter} C.~N., {Morganti} R., {Killeen} N.~E.~B., {Fosbury}
  R.~A.~E., {Hook} R.~N., {Siebert} J.,  {Shaw} M.~A., 1997, \mnras, 286, 558

\bibitem[\protect\citeauthoryear{{Conway}, {Burn}, \& {Vall{\'e}e}}{{Conway}
  et~al.}{1977}]{conway77}
{Conway} R.~G., {Burn} B.~J.,  {Vall{\'e}e} J.~P., 1977, \aaps, 27, 155

\bibitem[\protect\citeauthoryear{{Crawford} et~al.}{{Crawford}
  et~al.}{1999}]{crawford99}
{Crawford} C.~S., {Allen} S.~W., {Ebeling} H., {Edge} A.~C.,  {Fabian} A.~C.,
  1999, \mnras, 306, 857

\bibitem[\protect\citeauthoryear{{Crawford} \& {Fabian}}{{Crawford} \&
  {Fabian}}{1989}]{crawfordfabian89}
{Crawford} C.~S.,  {Fabian} A.~C., 1989, \mnras, 239, 219

\bibitem[\protect\citeauthoryear{{Crawford} et~al.}{{Crawford}
  et~al.}{2002}]{c02}
{Crawford} C.~S., {Gandhi} P., {Fabian} A.~C., {Wilman} R.~J., {Johnstone}
  R.~M., {Barger} A.~J.,  {Cowie} L.~L., 2002, \mnras, 333, 809

\bibitem[\protect\citeauthoryear{{Crawford} \& {Vanderriest}}{{Crawford} \&
  {Vanderriest}}{1997}]{crawfordvanderriest97}
{Crawford} C.~S.,  {Vanderriest} C., 1997, \mnras, 285, 580

\bibitem[\protect\citeauthoryear{{Della Ceca} et~al.}{{Della Ceca}
  et~al.}{2003}]{dellaceca03}
{Della Ceca} R. et~al., 2003, \aap, 406, 555

\bibitem[\protect\citeauthoryear{{Dopita} \& {Sutherland}}{{Dopita} \&
  {Sutherland}}{1995}]{dopitasutherland95}
{Dopita} M.~A.,  {Sutherland} R.~S., 1995, \apj, 455, 468

\bibitem[\protect\citeauthoryear{{Eales} et~al.}{{Eales}
  et~al.}{1993}]{eales93}
{Eales} S.~A., {Rawlings} S., {Dickinson} M., {Spinrad} H., {Hill} G.~J.,
  {Lacy} M., 1993, \apj, 409, 578

\bibitem[\protect\citeauthoryear{{Elvis} et~al.}{{Elvis}
  et~al.}{1994}]{elvis94}
{Elvis} M. et~al., 1994, \apjs, 95, 1

\bibitem[\protect\citeauthoryear{{Fabian}}{{Fabian}}{1999}]{f99}
{Fabian} A.~C., 1999, \mnras, 308, L39

\bibitem[\protect\citeauthoryear{{Fabian} et~al.}{{Fabian}
  et~al.}{2001}]{3c294mainref}
{Fabian} A.~C., {Crawford} C.~S., {Ettori} S.,  {Sanders} J.~S., 2001, \mnras,
  322, L11

\bibitem[\protect\citeauthoryear{{Fabian} et~al.}{{Fabian}
  et~al.}{1987}]{fabian87}
{Fabian} A.~C., {Crawford} C.~S., {Johnstone} R.~M.,  {Thomas} P.~A., 1987,
  \mnras, 228, 963

\bibitem[\protect\citeauthoryear{{Fabian}, {Wilman}, \& {Crawford}}{{Fabian}
  et~al.}{2002}]{fabianwilmancrawford02}
{Fabian} A.~C., {Wilman} R.~J.,  {Crawford} C.~S., 2002, \mnras, 329, L18

\bibitem[\protect\citeauthoryear{{Fanaroff} \& {Riley}}{{Fanaroff} \&
  {Riley}}{1974}]{fanaroff_riley}
{Fanaroff} B.~L.,  {Riley} J.~M., 1974, \mnras, 167, 31P

\bibitem[\protect\citeauthoryear{{Fanti} et~al.}{{Fanti}
  et~al.}{1995}]{fanti95}
{Fanti} C., {Fanti} R., {Dallacasa} D., {Schilizzi} R.~T., {Spencer} R.~E.,
  {Stanghellini} C., 1995, \aap, 302, 317

\bibitem[\protect\citeauthoryear{{Ferland} et~al.}{{Ferland}
  et~al.}{1998}]{cloudy0507}
{Ferland} G.~J., {Korista} K.~T., {Verner} D.~A., {Ferguson} J.~W., {Kingdon}
  J.~B.,  {Verner} E.~M., 1998, \pasp, 110, 761

\bibitem[\protect\citeauthoryear{{Freeman} et~al.}{{Freeman}
  et~al.}{2002}]{wavdetect}
{Freeman} P.~E., {Kashyap} V., {Rosner} R.,  {Lamb} D.~Q., 2002, \apjs, 138,
  185

\bibitem[\protect\citeauthoryear{{Gandhi} et~al.}{{Gandhi} et~al.}{2004}]{g04}
{Gandhi} P., {Crawford} C.~S., {Fabian} A.~C.,  {Johnstone} R.~M., 2004,
  \mnras, 348, 529

\bibitem[\protect\citeauthoryear{{Gardner} et~al.}{{Gardner}
  et~al.}{1997}]{gardner97}
{Gardner} J.~P., {Sharples} R.~M., {Frenk} C.~S.,  {Carrasco} B.~E., 1997,
  \apjl, 480, L99

\bibitem[\protect\citeauthoryear{{Grimes}, {Rawlings}, \& {Willott}}{{Grimes}
  et~al.}{2004}]{grimes04}
{Grimes} J.~A., {Rawlings} S.,  {Willott} C.~J., 2004, \mnras, 349, 503

\bibitem[\protect\citeauthoryear{{Iwasawa}}{{Iwasawa}}{1999}]{iwasawa99}
{Iwasawa} K., 1999, \mnras, 302, 96

\bibitem[\protect\citeauthoryear{{Iwasawa} et~al.}{{Iwasawa}
  et~al.}{2005}]{iwasawa05}
{Iwasawa} K., {Crawford} C.~S., {Fabian} A.~C.,  {Wilman} R.~J., 2005, \mnras,
  362, L20

\bibitem[\protect\citeauthoryear{{Iwasawa}, {Fabian}, \& {Ettori}}{{Iwasawa}
  et~al.}{2001}]{iras09mainref}
{Iwasawa} K., {Fabian} A.~C.,  {Ettori} S., 2001, \mnras, 321, L15

\bibitem[\protect\citeauthoryear{{Lavery}, {Owen}, \& {Henry}}{{Lavery}
  et~al.}{1993}]{lavery93}
{Lavery} R.~J., {Owen} F.~N.,  {Henry} J.~P., 1993, American Astronomical
  Society Meeting, 25, 1307

\bibitem[\protect\citeauthoryear{{Lilly} et~al.}{{Lilly}
  et~al.}{1995}]{lilly95}
{Lilly} S.~J., {Tresse} L., {Hammer} F., {Crampton} D.,  {Le Fevre} O., 1995,
  \apj, 455, 108

\bibitem[\protect\citeauthoryear{{Machalski} \& {Condon}}{{Machalski} \&
  {Condon}}{1983}]{machalskicondon83}
{Machalski} J.,  {Condon} J.~J., 1983, \aj, 88, 143

\bibitem[\protect\citeauthoryear{{Magdziarz} \& {Zdziarski}}{{Magdziarz} \&
  {Zdziarski}}{1995}]{pexrav}
{Magdziarz} P.,  {Zdziarski} A.~A., 1995, \mnras, 273, 837

\bibitem[\protect\citeauthoryear{{Mainieri} et~al.}{{Mainieri}
  et~al.}{2002}]{mainieri02}
{Mainieri} V., {Bergeron} J., {Hasinger} G., {Lehmann} I., {Rosati} P.,
  {Schmidt} M., {Szokoly} G.,  {Della Ceca} R., 2002, \aap, 393, 425

\bibitem[\protect\citeauthoryear{{Mainieri} et~al.}{{Mainieri}
  et~al.}{2005}]{mainieri05}
{Mainieri} V. et~al., 2005, \mnras, 356, 1571

\bibitem[\protect\citeauthoryear{{Maiolino} et~al.}{{Maiolino}
  et~al.}{2001}]{maiolino01}
{Maiolino} R., {Marconi} A., {Salvati} M., {Risaliti} G., {Severgnini} P.,
  {Oliva} E., {La Franca} F.,  {Vanzi} L., 2001, \aap, 365, 28

\bibitem[\protect\citeauthoryear{{Maiolino} et~al.}{{Maiolino}
  et~al.}{2006}]{maiolino06}
{Maiolino} R. et~al., 2006, \aap, 445, 457

\bibitem[\protect\citeauthoryear{{Mateos} et~al.}{{Mateos}
  et~al.}{2005}]{mateos05}
{Mateos} S. et~al., 2005, \aap, 433, 855

\bibitem[\protect\citeauthoryear{{McCarthy}, {van Breugel}, \&
  {Kapahi}}{{McCarthy} et~al.}{1991}]{mccarthy91}
{McCarthy} P.~J., {van Breugel} W.,  {Kapahi} V.~K., 1991, \apj, 371, 478

\bibitem[\protect\citeauthoryear{{McCarthy} et~al.}{{McCarthy}
  et~al.}{1987}]{mccarthy87}
{McCarthy} P.~J., {van Breugel} W., {Spinrad} H.,  {Djorgovski} S., 1987,
  \apjl, 321, L29

\bibitem[\protect\citeauthoryear{{McMahon} et~al.}{{McMahon}
  et~al.}{2002}]{mcmahon02}
{McMahon} R.~G., {White} R.~L., {Helfand} D.~J.,  {Becker} R.~H., 2002, \apjs,
  143, 1

\bibitem[\protect\citeauthoryear{{Mushotzky}, {Done}, \& {Pounds}}{{Mushotzky}
  et~al.}{1993}]{mushotzky93}
{Mushotzky} R.~F., {Done} C.,  {Pounds} K.~A., 1993, \araa, 31, 717

\bibitem[\protect\citeauthoryear{{Norman} et~al.}{{Norman}
  et~al.}{2002}]{norman02}
{Norman} C. et~al., 2002, \apj, 571, 218

\bibitem[\protect\citeauthoryear{{Owen}, {White}, \& {Ge}}{{Owen}
  et~al.}{1993}]{owen93}
{Owen} F.~N., {White} R.~A.,  {Ge} J., 1993, \apjs, 87, 135

\bibitem[\protect\citeauthoryear{{Padrielli} \& {Conway}}{{Padrielli} \&
  {Conway}}{1977}]{padrielliconway77}
{Padrielli} L.,  {Conway} R.~G., 1977, \aaps, 27, 171

\bibitem[\protect\citeauthoryear{{Pedelty} et~al.}{{Pedelty}
  et~al.}{1989}]{pedelty89b}
{Pedelty} J.~A., {Rudnick} L., {McCarthy} P.~J.,  {Spinrad} H., 1989, \aj, 98,
  1232

\bibitem[\protect\citeauthoryear{{Perola} et~al.}{{Perola}
  et~al.}{2004}]{perola04}
{Perola} G.~C. et~al., 2004, \aap, 421, 491

\bibitem[\protect\citeauthoryear{{Riley}}{{Riley}}{1975}]{riley75}
{Riley} J.~M., 1975, \mnras, 170, 53

\bibitem[\protect\citeauthoryear{{Saikia} et~al.}{{Saikia}
  et~al.}{2001}]{saikia01}
{Saikia} D.~J., {Jeyakumar} S., {Salter} C.~J., {Thomasson} P., {Spencer}
  R.~E.,  {Mantovani} F., 2001, \mnras, 321, 37

\bibitem[\protect\citeauthoryear{{Severgnini} et~al.}{{Severgnini}
  et~al.}{2006}]{severgnini06}
{Severgnini} P. et~al., 2006, A\&A accepted, astro-ph/0602486

\bibitem[\protect\citeauthoryear{{Stark} et~al.}{{Stark}
  et~al.}{1992}]{stark92}
{Stark} A.~A., {Gammie} C.~F., {Wilson} R.~W., {Bally} J., {Linke} R.~A.,
  {Heiles} C.,  {Hurwitz} M., 1992, \apjs, 79, 77

\bibitem[\protect\citeauthoryear{{Stern} et~al.}{{Stern}
  et~al.}{2002}]{stern02}
{Stern} D. et~al., 2002, \apj, 568, 71

\bibitem[\protect\citeauthoryear{{Stockton} \& {MacKenty}}{{Stockton} \&
  {MacKenty}}{1987}]{stocktonmackenty87}
{Stockton} A.,  {MacKenty} J.~W., 1987, \apj, 316, 584

\bibitem[\protect\citeauthoryear{{Sturm} et~al.}{{Sturm}
  et~al.}{2006}]{sturm06}
{Sturm} E., {Hasinger} G., {Lehmann} I., {Mainieri} V., {Genzel} R., {Lehnert}
  M.~D., {Lutz} D.,  {Tacconi} L.~J., 2006, ApJ accepted, astro-ph/0601204

\bibitem[\protect\citeauthoryear{{Tozzi} et~al.}{{Tozzi}
  et~al.}{2006}]{tozzi06}
{Tozzi} P. et~al., 2006, \aap, 451, 457

\bibitem[\protect\citeauthoryear{{Willis}, {Hewett}, \& {Warren}}{{Willis}
  et~al.}{2005}]{willis05b}
{Willis} J.~P., {Hewett} P.~C.,  {Warren} S.~J., 2005, \mnras, 363, 1369

\bibitem[\protect\citeauthoryear{{Willott} et~al.}{{Willott}
  et~al.}{2003}]{willott03}
{Willott} C.~J. et~al., 2003, \mnras, 339, 397

\bibitem[\protect\citeauthoryear{{Wills}}{{Wills}}{1976}]{wills76}
{Wills} B.~J., 1976, \aj, 81, 1031

\bibitem[\protect\citeauthoryear{{Wilman} et~al.}{{Wilman}
  et~al.}{2003}]{wilman03_hyperluminous}
{Wilman} R.~J., {Fabian} A.~C., {Crawford} C.~S.,  {Cutri} R.~M., 2003, \mnras,
  338, L19

\bibitem[\protect\citeauthoryear{{Young} et~al.}{{Young}
  et~al.}{2002}]{young02}
{Young} A.~J., {Wilson} A.~S., {Terashima} Y., {Arnaud} K.~A.,  {Smith} D.~A.,
  2002, \apj, 564, 176

\bibitem[\protect\citeauthoryear{{Zakamska} et~al.}{{Zakamska}
  et~al.}{2004}]{zakamska04_II}
{Zakamska} N.~L., {Strauss} M.~A., {Heckman} T.~M., {Ivezi{\' c}} {\v Z}.,
  {Krolik} J.~H., 2004, \aj, 128, 1002

\bibitem[\protect\citeauthoryear{{Zamorani} et~al.}{{Zamorani}
  et~al.}{1981}]{zamorani81}
{Zamorani} G. et~al., 1981, \apj, 245, 357

\end{thebibliography}

\end{document}